\documentclass[a4paper]{article}

\usepackage[usenames,dvipsnames,svgnames]{xcolor}

\usepackage[english]{babel}
\usepackage[utf8]{inputenc}
\usepackage{amsmath}
\usepackage{graphicx}
\usepackage[colorinlistoftodos]{todonotes}
\usepackage{import}

\usepackage{hyperref}
\hypersetup{colorlinks=true,linkcolor=blue,citecolor=red}

\usepackage{my-own-commands}

\title{Geometrical  Aspects of non-gravitational interactions}

\author{Omar Roldan$^{1,2}$, C.C. Barros Jr. $^1$\\
$^1$Universidade Federal de Santa Catarina, $^2$Universidade Federal do Rio de Janeiro
}

\date{}

\usepackage[a4paper]{geometry}

\begin{document}

\maketitle

\begin{abstract}
In this work we look for a geometric description of non-gravitational forces. The basic ideas are proposed studying the interaction between a punctual particle and an electromagnetic external field. For this purpose, we introduce the concept of proper space-time, that allow us to describe this interaction in a way analogous to the one that the general relativity theory does for gravitation. The field equations that define this geometry are similar to the Einstein's equations, where in general, the energy-momentum tensor have information of both, the particle and the external field. In this formalism we consider the particle path as being a geodesic in a curved space-time, and so, the electromagnetic force is understood in a purely geometric way.
\end{abstract}


\section*{Introduction}

As our knowledge of Physics improves, more and more the general theory of relativity is showing to be a fundamental theory.  Among many important results that could be  listed, the recent  detection of gravitational waves by the LIGO experiment \cite{ligo} is just one example.

Although, when considering non-gravitational interactions  there are still some questions to be answered. An interesting aspect to be observed is the preferential role  of the gravitational interactions in this theory, as all interactions are coupled to the gravity by the gravitational constant {\it G} in the right-hand side of the Einstein equations. A fundamental question is if the gravitational interaction really is a preferential kind of interaction or if all interactions may be considered in an equivalent way. Another question that may be posed is how a system where the gravity is negligible if compared with the other kinds of interactions may be considered in a geometrical approach.  Some interesting results of these ideas at the quantum level  have been shown in \cite{Barros:2004ta} and \cite{Barros:2005jj}. In this work, this possibility will be studied and the formulation of the basic ideas will be made in terms of the interaction of a point-like particle with an external electromagnetic field.

The theory of General Relativity  (GR) describes the gravitation as a geometrical effect. In this way, in absence of non-gravitational forces\footnote{In this work, our main interest is in studying interactions other than gravitation.}, any particle will follow a geodesic path on a given curved space-time (hereafter s-t) as described by the geodesic equation
\be \label{geod}
\derr{x^\mu }{\tau} + \Gmrs\der{x^\rho }{\tau}\der{x^\sigma }{\tau} =0\,.
\ee 
%

Being a geometrical effect, it implies that the evolution of each particle in a gravitational field is independent on their characteristics such as charge, mass or spin.

A different situation is the case of a charged particle interacting with an electromagnetic field (hereafter \Em), in which the trajectory is determined by
\be \label{force}
\derr{x^\mu }{\tau} + \Gmrs\der{x^\rho }{\tau}\der{x^\sigma}{\tau} = \f{1}{m} f^\mu,
\ee 
where $m$ is the mass of the particle and $f^\mu$ is the Lorentz force which depends on the particle's charge. Then, when introducing interactions other than gravitation in the usual formulation of GR, the dynamics of particles is not (completely) described in geometrical terms.

In this work, we want to \textit{effectively} describe the electromagnetic interaction as a geometrical phenomena. That is, we begin with the assumption that the path followed by a charged particle in presence of an \Em, correspond to a geodesic in a certain curved s-t, which we will call the particle's proper space-time (\pst). Thinking in this way, each particle is in free fall in its own s-t, which must necessarily depend on the interaction between the particle and its environment (the external fields).

Then, we must carefully distinguish between this abstract s-t (the \pst) and the s-t which is perceived by the observer in the laboratory (our s-t). We will call this last one the background space-time (\bst).

To our knowledge, there is no theory at present which totally describes the electromagnetic interaction in a geometric way. Many people have care about this issue, moved for example by Einstein's spirit which in its own words is: ``The idea that there are two structures of space independent of each other, the metric-gravitational and the electromagnetic, is intolerable to the theoretical spirit" (Aspden \cite{aspden}, and Barnett \cite{Barnett}).

Hermann Weyl (Pauli \cite{Pauli}, see also \cite{Weyl:1929fm}), developed a generalized version of Riemann geometry to describe the electromagnetism in a geometrical way. Even though it was a very deep theory in which he has able to obtain the Maxwell equations by just using geometry, this formalism was practically abandoned, as it predicts some effects which are not consistent with observations (Pauli \cite{Pauli})

It's not the purpose of this work to give a geometric description of electromagnetism, but just to gain some insights on how to proceed in order to effectively describe the interaction of a charged particle and an \Em. Some useful results can be obtained with this approach as it was done for example, in the work of Barros \cite{Barros:2004ta,Barros:2005jj,Barros:2005mh}, in which by  using the spherical symmetry in the interaction of electron with a proton in the Hydrogen atom, he managed to obtain a Minkowski-like metric and it allows him to obtain the spectrum of the system which is very close to the observed one.

This work is organized in four sections. In section 1. we  make the weak field approximation and the limit of small velocity to get a first relationship between the metric of the \pst, the external \Em\ and the particle mass and charge. Section 2. sets the main ideas on how to obtain the metric for the \pst\ in more general cases. Section 3 and 4. give the applications of the present formalism and its comparison with the results of the Special Relativity (SR). In the last section the conclusions will be drawn.

%
\section{Weak field and small velocity limit} \label{sec:weak}

\subsection{Geodesic equation and Lorentz force} \label{lorentz}

The first step in order to construct the \pst\ is to determine it in the weak field limit and its relation with a flat s-t, in a way similar to the one that is done in the General Relativity Theory. \\

In the weak field approximation we can think of the \pst\ as described by a metric $g_{\Mn}$ which is a small perturbation of the \bst. For simplicity we shall take this \bst\ to be flat and so described by the Minkowski metric%
\footnote{This approach is common in perturbation theory for the study of gravitational waves, see: \cite{Carroll:2004st,Weinberg:1972kfs,Wald:1984rg,deFelice:1990hu}.} $\eta_{\Mn}$
\be  \label{g}
g_{\Mn}=\eta_{\Mn}+h_{\Mn}, \quad |h_{\mu \nu}|\ll 1.
\ee

Here $h_{\Mn}$ is understood as a two-tensor in the \bst, and is responsible for the force on the charged particle (as perceived by the observer). The inverse metric at linear order in $h$ is
\be
g^{\Mn}=\eta^{\Mn}-h^{\Mn}, \qquad  h^{\Ml}=\eta^{\Mn}\eta^{\Ls} h_{\Ns},
\ee
that is, we rise and lower indices with the background metric.

The \cris \ $\Gmrs$ for the metric $g$ are expressed into the usual form
\be \label{crist}
\Gmrs=\f12 g^{\Ml}(-g_{\Rs,\lambda}+g_{\Lr,\Os}+g_{\Sl,\rho}),
\ee
which can be related to the \cris \ $\TGmrs$ of the background metric $\eta$  by%
\footnote{In Cartesian coordinates $\TGmrs$ vanishes, but this is not the case in spherical coordinates for example.  Using $\TGmrs$ explicitly allows us to easily generalize our present treatment to a curved \bst.}
\begin{align}\label{simbolos}
\Gmrs &=\f12 (\eta^{\Ml}-h^{\Ml}) \Big[-(\eta_{\Rs,\lambda}+h_{\Rs,\lambda})+(\eta_{\Lr,\Os}+h_{\Lr,\Os})+(\eta_{\Sl,\rho}+h_{\Sl,\rho}) \Big]\notag \\
&= \TGmrs-\Amrs\,,
\end{align}
where we introduced
\begin{align}
\Amrs \equiv \frac12 h^{\Ml}\Big[\eta_{\Rs,\lambda}+\eta_{\Lr,\Os}+\eta_{\Sl,\rho} \Big]-\frac12 \eta^{\Ml}\Big[-h_{\Rs,\lambda}+h_{\Lr,\Os}+h_{\Sl,\rho} \Big]\,.
\end{align}

Unlike $\TGmrs$, $\Amrs$ is a tensor, as we now see. To prove this, it is easier to consider $\amrs$ instead of $\Amrs$, 
\begin{align}\label{a}
\amrs=&-\frac12 (-h_{\Rs,\mu}+h_{\Mr,\Os}+h_{\Sm,\rho})+\tilde{\Gamma}^\lambda_{\Rs}h_{\Ml}\notag\\
=&-\frac12 (-\nabla_\mu h_{\Rs}+\nabla_\Os h_{\Mr}+\nabla_\rho h_{\Sm}),
\end{align}
where $\nabla$ means the covariant derivative with respect to the \bst, that is
\be
\nabla_\mu h_{\Rs}=\p_\mu h_{\Rs}-\tilde{\OG}^\lambda_{\Mr}h_{\Ls}-\tilde{\Gamma}^\lambda_{\Ms}h_{\Rl}\,,
\ee
and the statement about $\amrs$ is verified. Now the main idea of the \pst\ concept, is that the particle will follow a geodesic in this curved s-t, so the particle's path should obey 
\be
\f{d^2 x^\mu }{d\tau^2} + \Gmrs\f{dx^\rho }{d\tau}\f{dx^\sigma }{d\tau}=0,
\ee
or
\begin{align} \label{force1}
\frac{d^2x^\mu }{d\tau^2} + \TGmrs\frac{dx^\rho }{d\tau}\frac{dx^\sigma }{d\tau}=\Amrs\frac{dx^\rho }{d\tau}\frac{dx^\sigma }{d\tau}.
\end{align}

If we take the limit of small velocities the proper time $\tau$ in the \bst \ and in the \pst\ are the same, and we can set $\tau=t$, where $t$ is the time as measured by the observer. In addition\footnote{We are using $x^0=ct$.}, $dx^i/d\tau\ll dx^0/d\tau \approx c$, then by comparing \eq{force1} with \eq{force} we can say that the observer in the laboratory will measure a force $f^\mu$ given by \footnote{Note that we are only considering the spatial component $i$ of the force, as the temporal component would require to consider $t\ne \tau$.}
\be \label{force2}
f^i=m A^i_{\ \Rs} \frac{dx^\rho }{d\tau}\frac{dx^\sigma}{d\tau} =m \lc A^i_{\ 00} \der{x^0 }{\tau} \der{x^0 }{\tau}+
2A^i_{\ 0j} \der{x^0 }{\tau} \der{x^j }{\tau}\rc\,,
\ee
where we used the fact that $\Amrs$ is symmetric in $\Rs$.

To continue, let's stress that we are considering an inertial observer in a flat \bst, whose metric in Cartesian coordinates is
\be
\eta_{\mu \nu}= \textrm{diag}(-1,1,1,1).
\ee

Now, if this observer choose any other coordinate system to describe the spatial part, say, spherical coordinates, but without changing the temporal part, then%
\be
\eta_{\mu \nu}=
\begin{pmatrix}
-1 & 0 & 0 & 0 \\
0 & \eta_{11} & \eta_{12} & \eta_{13} \\
0 & \eta_{21} & \eta_{22} & \eta_{23} \\
0 & \eta_{31} & \eta_{32} & \eta_{33} 
\end{pmatrix},
\ee 
where in general $\eta_{ij}=\eta_{ij}(\vec{x})\ne 0$ for $i\ne j$. Note that if we make a boost this will affect the time component, but as far as we take the limit of small velocities, we are somehow performing a Galilean transformation of coordinates, which does not affect the time component of the metric.

Considering this convention we have that in general $\TGmrs\ne 0$, but%
\be
\tilde{\Gamma}^0_{\Ms}=\tilde{\Gamma}^\lambda_{\mu 0}=\tilde{\Gamma}^\lambda_{0 \Os}=0.
\ee

Using this fact we can easily compute
\be \begin{split} \label{cova}
&\nabla_0 h_{\Rs}=\p_0 h_{\Rs}, \quad \nabla_i h_{00}=\p_i h_{00},\\
&\nabla_i h_{0j}=\p_i h_{0j}-\tilde{\OG}^k_{ij} h_{0k},
\end{split}
\ee
so that
\be \label{Aijk1}
A_{i 0 0}=\p_i \f{h_{00}}{2}-\p_0 h_{i0}, \quad A_{i j 0}=\f12 \lp \p_i h_{j0}-\p_j h_{i0}\rp -\f12 \p_0 h_{ij}.
\ee

If we relate $h_{\mu 0}$ with the electromagnetic potential through 
\be \label{ha}
h_{0 0}=2\f{q}{mc} A_{0}, \quad h_{0 i}=\f{q}{mc} A_{i}\,,
\ee
we see that \eq{Aijk1} yields
\begin{align} 
&A_{i 0 0}= \f{q}{mc} F_{i0},  \quad A_{i j 0}=\f{q}{2mc} F_{ij}-\f12 \p_0 h_{ij},\label{af}
\end{align}
where $F_{\Mn}=\p_\mu A_\nu-\p_\nu A_\mu$ is the EM tensor. Substituting \eq{af} into (\ref{force2}) give us 
\begin{align}
f^i=q \lc c F^i_{\ 0} + F^i_{\ j}v^j \rc-q\ \p_0 h^i_{\ j} v^j \,,
\end{align}
which is the Lorentz force
$f^i=q F^i_{\ \rho} dx^\rho/ d\tau$ plus an additional term, $q\ \p_0 h^i_{\ j}$, which we expect to be subdominant due to the time derivative $\p_0=\f1 c\p_t$ (at least for slowly time-varying \Em s) and vanishes for the cases of interest in the present paper: static \Em s. The more general situation will be addressed in a future paper.

We want to stress that the \Emp\ $A_\mu$ is a vector in the \bst, and by construction $h_{\Mn}$ is a tensor. Then, it seems that the quantity $(h_{00}/2,h_{0i})$ cannot be identified with a vector as we suggested in \eq{ha}. However, as it was said, in this paper we only allow spatial coordinates transformations (or Galilean transformations)%
\footnote{E.g., we can go from Cartesian $\lp t,x,y,z\rp$ to spherical coordinates $\lp t,r,\theta,\phi\rp$.}
\be
x^{\mu}=(x^0,x^k)\to x^{\mu'}=(x^0,x^{k'}),
\ee
and so, $A_\mu$ transform as
\be
A_{\mu'}=\f{\p x^\mu}{\p x^{\mu'}}A_\mu=\lp A_0,\ \f{\p x^{k'}}{\p x^k} A_k\rp.
\ee

The transformation for $h_{0\mu}$ is given by
\begin{align}
h_{0\mu'}=\f{\p x^\nu}{\p x^0} \f{\p x^\mu}{\p x^{\mu'}} h_{\Nm}=\f{\p x^\mu}{\p x^{\mu'}} h_{0\mu}\notag \\
\to \quad h_{0\mu'}=\lp h_{00},\ \f{\p x^k}{\p x^{k'}} h_{0k}\rp,
\end{align}
that is, $h_{0\mu}$ has the same behavior as $A_\mu$ under this restricted group of transformations%
\footnote{We can also note that this property applies to the covariant derivative. That is, from \eq{cova} we see that $h_{00}$ behaves like a scalar while $h_{0i}$ transform like a vector.}. %
We leave the discussion about a possible generalization to any group of transformations for a future paper.


\subsection{Riemann Tensor and gauge freedom}

The Riemann tensor for the \pst\ is
\be \label{riem}
R^\mu_{\ \ \Nr \Os}=-\Gamma^{\mu}_{\Nr, \Os}+\Gamma^{\mu}_{\Sn, \rho}%
-\Gamma^{\lambda}_{\Nr}\Gamma^{\mu}_{\Sl}+\Gamma^{\lambda}_{\Sn}\Gamma^{\mu}_{\Rl},
\ee
which can be split as
\begin{align}
&\Rmnrs=-\lp \tilde\OG^{\mu}_{\Nr, \Os}-A^{\mu}_{\ \ \Nr, \Os}\rp+%
\lp \tilde\OG^{\mu}_{\Sn, \rho}-A^{\mu}_{\ \ \Sn, \rho}\rp\notag \\%
&-\lp \tilde\OG^{\lambda}_{\Nr}-A^{\lambda}_{\ \ \Nr}\rp \lp \tilde\OG^{\mu}_{\Sl}-A^{\mu}_{\ \ \Sl}\rp %
+\lp \tilde\OG^{\lambda}_{\Sn}-A^{\lambda}_{\ \ \Sn}\rp \lp \tilde\OG^{\mu}_{\Rl}-A^{\mu}_{\ \ \Rl}\rp. \notag
\end{align}

In first order in $h$ (and in its derivatives) we can write $\Rmnrs=(R_0+R_1)^{\mu}_{\ \ \nu \rho \sigma}$, where $R_0$ refers to the Riemann tensor in the \bst, so for our case $(R_0)^{\mu}_{\ \ \nu \rho \sigma}=0$ (flat s-t). Then,
\begin{align}
\Rmnrs=&(R_1)^{\mu}_{\ \ \nu \rho \sigma}=%
A^{\mu}_{\ \ \Nr, \Os}-A^{\mu}_{\ \ \Sn, \rho}+\tilde\Gamma^{\lambda}_{\Nr}A^{\mu}_{\ \ \Sl}\notag \\%
&+A^{\lambda}_{\ \ \Nr}\tilde\Gamma^{\mu}_{\Sl}%
-\tilde\Gamma^{\lambda}_{\Sn}A^{\mu}_{\ \ \Rl}-A^{\lambda}_{\ \ \Sn}\tilde\Gamma^{\mu}_{\Rl},
\end{align}
which can be written in a compact form as
\be \label{riem1}
\Rmnrs=\nabla_\Os A^{\mu}_{\ \ \Nr}-\nabla_\rho A^{\mu}_{\ \ \Sn}.
\ee

From \eq{force1} we see that the quantity we can directly measure is $\Amrs$, once this is directly related to the force $f^\mu$ or to the particle's path. Then, in accordance with \eq{a} we can think of  $h_{\Ms}$ as an interacting potential. This leads necessarily to a gauge freedom for the tensor $h_{\Ms}$ and so for the metric $g_{\Ms}$.

We can then ask, which is the more general ``group" of transformations on $h_{\Mn}$ that leaves the dynamics of the particle invariant as perceived by the observer. To start with, we can seek for which transformations leave the Riemann  tensor invariant. This was already solved  (see e.g., Carroll \cite{Carroll:2004st})
\be \label{gauge}
h_{\Mn}\to h_{\Mn}+\nabla_\mu \xi_\nu+\nabla_\nu \xi_\mu,
\ee
where $\xi_\mu$ is an arbitrary vector%
\footnote{The only restriction on $\xi$ is that is small enough such that the condition $|h_\Mn|\ll 1$ is satisfied.}%
in the \bst. However, $\Amrs$ is not invariant under this general transformation as we will see now. From \eq{a} we have
\be \begin{split}
\amrs\to&\frac12 \Bigg\{-\nabla_\mu \lp h_{\Rs}+\nabla_\rho \xi_\Os+\nabla_\Os \xi_\rho \rp \notag\\%
&+\nabla_\Os \lp h_{\Mr}+\nabla_\mu \xi_\rho+\nabla_\rho \xi_\mu \rp 
+\nabla_\rho \lp h_{\Sm}+\nabla_\Os \xi_\mu+\nabla_\mu \xi_\Os \rp \Bigg\},
\end{split}
\ee
which leads to
\be \label{trans-A}
\amrs \to \amrs-\nabla_\rho \nabla_\Os \xi_\mu,
\ee
and we have used $\nabla_\rho \nabla_\Os=\nabla_\Os \nabla_\rho$ which is valid for a flat s-t, as we assumed to be the case for the \bst.

Using this transformation we can easily see that the Riemann tensor is invariant. Substituting into \eq{riem1} we get
\be \begin{split}
R_{\Mn\Rs}\to & \quad \nabla_\Os \lp A_{\mu\Nr}-\nabla_\nu\nabla_\rho \xi_\mu \rp%
-\nabla_\rho \lp A_{\mu\Sn}-\nabla_\Os\nabla_\nu \xi_\mu \rp \\
&=R_{\Mn\Rs}.
\end{split}
\ee

Though $\amrs$ is not invariant under the full group of transformations (\eq{gauge}), if we fix the three spatial components $\xi^i$, then the gauge freedom is due only to $\xi^0$ and we get the restricted transformation
\be \label{gauge1}
h_{00}\to h_{00}+2\nabla_0 \xi_0, \quad h_{0i}\to h_{0i}+\nabla_i \xi_0, \quad h_{ij}\to h_{ij}.
\ee

Defining $\xi_0=\lp q/mc\rp \phi$, we have the gauge transformation
\be \label{gauge1}
\lp \f{h_{00}}{2},h_{0i}\rp \to \lp \f{h_{00}}{2}+\nabla_0\phi, h_{0i}+\nabla_i\phi\rp\,,
\ee
which is perfectly consistent with the gauge transformation $A_\mu\to A_\mu+\nabla_\mu \phi$, see \eq{ha}.


Under this restricted group of transformations the tensor $A$ transform according to (see \eq{trans-A})
\be \label{trans-A1}
A_{0\Rs} \to A_{0\Rs}-\nabla_\rho \nabla_\Os \xi_0, \quad A_{i\Rs} \to A_{i\Rs}\,,
\ee
and since $A_{i\Rs}$ is invariant, then the force $f_i$  (see \eq{force2})
\be
f_i=m A_{i\Rs}\frac{dx^\rho }{d\tau}\frac{dx^\sigma }{d\tau},
\ee
will also be invariant. However the temporal component 
\be
f^0=c\f{dp^0}{d\tau}=mc A^0_{\ \Rs}\frac{dx^\rho }{d\tau}\frac{dx^\sigma }{d\tau}\,,
\ee
which represents the energy transfer is not invariant. We leave the solution of this problem for a future paper.

In section \ref{sec:weak}, we used the limit of weak field and small velocities to obtain a relationship between the \pst\ metric and the external \Em. In the next section we go an step further and to try to tackle the more general case in which the fields and velocities do not need be small.


%
\section{Electromagnetic field in the \pst}\label{ch:campoFmn}

In Special Relativity and in Cartesian coordinates, the electromagnetic tensor  $F^{\Mn}$ is totally defined by  the electric and magnetic fields $\vec E$, $\vec B$ (See \cite{Carroll:2004st,eletro1,Jackson:1998nia,Landau:1982dva}) 
\be
F^{0i}\equiv E^i/c, \quad F_{ij}\equiv \tilde\epsilon_{ijk}B^k,
\ee
where $\tilde\epsilon_{ijk}$ is the totally antisymmetric Levi-Civita symbol, with $\tilde\epsilon_{123}=1$. However, in the general case, the tensor $F^{\Mn}$ is not completely defined by $\vec E$ and $\vec B$, since we need to know the metric (in this case, $\eta_{\Mn}$) in order to know $F_{0i}$ and $F^{ij}$. Let's study this in more detail.

In a curved s-t we need to use the Levi-Civita tensor, which in four dimensions is given by (see e.g  Carroll \cite{Carroll:2004st})
\be
\epsilon_{\Mn \Rs}\equiv \ \sqrt{-g} \tilde\epsilon_{\Mn \Rs}, \quad g=\textrm{det}(g_{\Mn}),
\ee
and once more $\tilde\epsilon_{\Mn \Rs}$ is totally antisymmetric with $\tilde\epsilon_{0123}=1$. The Levi-Civita tensor with upper indexes is
\be
\epsilon^{\Mn \Rs}\equiv \ \frac{\tilde\epsilon^{\Mn \Rs}}{\sqrt{-g}}, \quad \tilde\epsilon^{ 0123}=-1.
\ee

Then, the electric field determine the $F^{0i}$  components, while magnetic field determine (except by the factor $\sqrt{-g}$) the other components $F_{ij}$. So, in general
\be \label{eq. max}
F^{0i}\equiv E^i/c, \quad F_{ij}\equiv \epsilon_{0ijk}B^k.
\ee

Nonetheless, a complete knowledge of the EM tensor is only possible with the help of the metric, e.g., to know $F_{0i}=g_{0\mu}g_{i\nu} F^{\Mn}$ we need not only the metric $g$ as well as the components $F^{ij}$, which also depend on $F_{\Mn}$ by $F^{ij}=g^{i\mu}g^{j\nu} F_{\Mn}$.

Then, we see that in general the EM tensor $F_{\Mn}$ will depend on the particle's properties (because of the dependence on $g_{\Mn}$) but since the electric and magnetic fields $\vec{E}$ e $\vec{B}$ are external fields 
\be \label{EB}
E^i=c F^{0i}, \quad \text{and} \quad B^i\equiv \f12\epsilon^{0ijk} F_{jk},
\ee
they are independent on the particle's charge and mass. {\it $\vec E$ and $\vec B$ are not unknown quantities, they are imposed as external conditions.}

To proceed, we shall give three essential statements for a general \pst: \\

The first one is that for vanishing external \Em\ ($\vec{E}=\vec{B}=0$) the \pst\ must coincide with the \bst. \\

The second one is that, from the point of view of the \pst\ formalism, the charged particle follows a geodesic in a curved s-t which is determined by the interaction with the external EM fields $\vec{E}$ and $\vec{B}$, so it is natural that the EM tensor built through \eq{EB} must satisfy the Maxwell equations in the \pst. That is,
\be 
\nabla_\mu F^{\Nm}=\mu_0 J^\nu, 
\ee
where hereafter $\nabla_\mu$ is the covariant derivative in the \pst, and $J^\nu$ is the current density which sources the \Em. In addition, in the present work we are interested in the case in which the charged particle is moving in the vacuum (not in a medium) so we can set $J^\nu=0$. Note that in the \bst\ $J^\nu$ need not be zero everywhere, but it is only non-zero in places where our particle of interest in not moving on.\\

And as a third statement, we will consider that if we apply a force $f^i$ to this particle%
\footnote{Here, by force we mean any other interaction that is not explicitly include in the geometry of the \pst. It can be or not of EM type.} %
the equation of motion should be
\be
\derr{x^i }{\tau}+ \Gamma^i_{\Rs}\der{x^\rho }{\tau}\der{x^\sigma }{\tau} =\frac{f^i}{m}\,.
\ee 

Now, if we take $f^i$ such that the particle stays at rest, then
\be
0=\derr{x^i }{\tau} =-\Gamma^i_{00}\der{x^0 }{\tau}\der{x^0 }{\tau} +\frac{f^i}{m},
\ee 
or
\be
0=\der{p^i }{\tau} =-\Gamma^i_{00}\frac{(p^0)^2}{m} + f^i\,.
\ee 

As seen ``by the particle", this external force is equivalent to apply a Lorentz force in the opposite direction to the electric field which enters the \pst\ metric, then in general
\be 
f^i=-q F^i_{\ 0} \f{p^0}{m},\qquad \to \qquad \Gamma^i_{00}\ p^0=-q\ F^{i \mu} g_{ \mu 0}\,.
\ee

Using $p^{\mu}p_{\mu}=-m^2 c^2$, we got for the particle at rest
\be
m^2 c^2=-p^0 p_0=-g_{00}(p^0)^2 \quad \to \quad p^0=m c/\sqrt{-g_{00}}\,,
\ee
and then,
\be \label{gi00}
\Gamma^i_{00}=\frac{q}{mc}\ F^{ \mu i}g_{ \mu 0}\sqrt{-g_{00}}\,.
\ee

%

\section{Applications} \label{sec:appl}

In the previous section we have set the basis for obtaining the metric of the \pst. In the remaining of this work we will apply such principles to obtain the \pst\ metric defined in two cases of interest: a spherically symmetric electric field and a constant electric field.

\subsection{Spherically Symmetric \pst}

As a first application we will consider a particle of charge $q$ in the presence of an external electric field with spherical symmetry. We assume that exist coordinates $t,r,\Ot,\phi$ such that the metric is diagonal
\be \label{esferica}
ds^2=-e^a \lp c^2 dt^2\rp+e^b dr^2+r^2 d\theta^2+r^2 \sin^2 \theta d\phi^2,
\ee
with $a$ and $b$ functions of $r$ alone.

Since we do not know which are the field equations that would allow us to obtain the metric, we can try to gain some information about $g_\Mn$ by trying to solve the Maxwell equations
\be 
\nabla_\mu F^{\Nm}=\mu_0 J^\nu=0, 
\ee
whose only non-trivial solution is 
\be \label{r-corriente}
\nabla_\mu F^{0\mu}=\partial_r F^{0r}+\Gamma^\mu_{\mu r}F^{0r}=0.
\ee

From the \cris\ definition
\be 
\Gmrs=\f12 g^{\Ml}\lp-g_{\Rs,\lambda}+g_{\Lr,\Os}+g_{\Sl,\rho}\rp,
\ee
we obtain the following relation valid for a diagonal metric 
\be
\Gamma^\mu_{\mu r}=\f12 g^{\mu\mu}(g_{\mu\mu,r})=\f12 \p_r \sum_\mu \ln |g_{\mu\mu}|=\f12 \partial_r \ln(-g).
\ee

Therefore, substituting into \eq{r-corriente} yields
\be \label{eq. maxwell}
\partial_r \ln(-gE^2)=0, \quad g=-e^{(a+b)}r^4 \sin^2 \theta.
\ee

Since, $E=c F^{0r}$ is the electric field which is independent on the \pst\ as discussed in the previous section, it is totally described by the \bst\ and must have the form
\be 
E=\f{K}{r^2}, \qquad K=\f{Q}{4\pi \Oe_0},
\ee
for some constant source charge $Q$. Using this into \eq{eq. maxwell} we note that $a=-b$ or 
\begin{align}\label{a-b}
g^{rr}=-g_{00}\,,
\end{align}
which leads to
\be
\Gamma^r_{00}=-\f12 g^{rr}g_{00}'=\f12 g_{00}\ g_{00}'\,.
\ee

By using the general equation \eq{gi00}, we arrive at
\be
g_{00}'=2\f{qE}{mc^2} \sqrt{-g_{00}}\,,
\ee
whose solution is
\be \label{g00}
\sqrt{-g_{00}}=1+\f{q}{mc^2} K\lp \f{1}{r}-\f{1}{r_0}\rp\,.
\ee

Here, $r_0\ne 0$ is some constant which expresses the gauge freedom. In solving this equation the integration constant was chosen in such a way that $g_{00}=-1$ when there's no electric field, $K=0$. This result was first obtained by Barros \cite{Barros:2005mh}.

Let's see that this result is in agreement with the linear approximation as it should be. In the linear approximation  $g_{00}=\eta_{00}+h_{00}$ with $h_{00}=2\f{q}{m_0c}A_0$, where $A_0$ is the electrostatic potential. On the other hand, in the \bst\  $\f{E^r}{c}=F_{r0}=\partial_r A_0$, then
\be
A_0=- \frac{K}{c}\lp \f{1}{r}-\f{1}{r_0}\rp \qquad \to \qquad h_{00}=-2\f{q}{m_0c^2}K\lp \f{1}{r}-\f{1}{r_0}\rp\,,
\ee
which is exactly the same expression we obtain by expanding $g_{00}$ in \eq{g00}.

%

\subsection{Field equations}

Now that we do know which is the metric for the spherically symmetric case, we can try to find out a field equation for the \pst \ which allows us to obtain the metric in any case. We propose to search for an Einstein-like equation
\be
G_{\Mn}=k T_{\Mn},
\ee
for some constant $k$, and for a given \Emt\ $T_{\Mn}$ which should depend on the interaction of the particle with the \Em.

By using the metric of the previous section, and fixing the gauge, $r_0=\infty$, we arrive at
%
\begin{align}
G^\theta_{\ \theta}=G^\phi_{\ \phi}=-G^0_{\ 0}=-G^r_{\ r}= \lp \f{q}{mc^2}\rp^2 E^2\,.\label{ricci-esferico}
\end{align}

Note that, in this case the Einstein tensor is traceless, which is also the case for the EM \Emt \ 
\be \label{em-Tensor}
T^\mu_{\ \nu}=\f{1}{\mu_0} \lc-F^\mu_{\ \lambda}F^\lambda_{\ \nu}+\f14 \delta^\mu_{\ \nu} F^\alpha_{\ \beta}F^\beta_{\ \alpha}\rc,
\ee
so, we can try to use this as the source for our field equations. We need however to stress that this is a gauge dependent statement, as \eq{ricci-esferico} was obtained in a fixed gauge, and it can be shown that for a general $r_0$, the metric in \eq{g00} leads to
\begin{align}
G^\mu_{\ \mu}=\frac{2 \Oa  (2-\Oa)}{r^2}, \qquad \Oa=\f{q}{mc^2}\frac{K}{r_0}\,,
\end{align}
that is, the trace of the Einstein tensor does not vanishes.

In addition, the EM tensor in \eq{em-Tensor} should be treated as the EM tensor in the \pst\ not on the \bst, then indexes should be raised and lowered with the metric $g_\Mn$ not $\eta_\Mn$. 
Computing the \eq{em-Tensor} for our case, we have
\begin{align} \label{mom-ener-esferico}
T^\theta_{\ \theta}=T^\phi_{\ \phi}=-T^0_{\ 0}=-T^r_{\ r}=\f12 \Oe_0 E^2\,,
\end{align}
where we used $E=cF^{0r}$ and $\mu_0\ \Oe_0=1/c^2$. Then, by comparing \eq{mom-ener-esferico} with \eq{ricci-esferico} we have
\begin{align}\label{eint-eq1}
G_{\Mn}=\frac{2}{\Oe_0} \lp\frac{q}{mc^2}\rp^2 T_{\Mn}\,.
\end{align}

This result was first obtained by Barros \cite{Barros:2005mh}. Now that we are in possession of the field equations let's obtain the metric for the case of a constant electric field.

%

\subsection{Uniform Electric Field}\label{ch:2-uni}

Let's consider an electric field in the $z$-axis, which is independent of $t,x,y$. If we assume that the line element is diagonal in this coordinate system, then
\be \label{eixo z}
ds^2=-e^a \lp c^2 dt^2\rp+e^b(dx^2+dy^2)+e^d dz^2,
\ee
where $a,b$ and $c$ are functions of $z$ only. By symmetry reasons we choose $g_{xx}=g_{yy}$.

The Maxwell equations leads to
\be
\nabla_\mu F^{0\mu}=\partial_z F^{0z}+\Gamma^\mu_{\mu z}F^{0z}=0, 
\ee
or
\be 
\ln(e^a e^{2b} e^d E^2)=constant.
\ee

If now we set $E=E_0=constant$, then, $a+2b+d=constant$, and using the fact that $a=b=d=0$ for $E_0=0$, then%
\footnote{Also, this constant can be absorbed into $dt^2$.} 
\be \label{eq. z1}
d=-a-2b.
\ee

With this result, we may compute the Einstein tensor and the \Emt, whose non-vanishing components are
\begin{align}
G^0_{\ 0}&=W\left(a''+2 a' b'+a'^2 \rp, \\
G^z_{\ z}&=W\left(a''+2 a' b'+a'^2+2 b''+ 3b'^2 \right),\\
G^x_{\ x}&=G^y_{\ y}=W\left(b''+a' b'+2b'^2 \rp \label{gxx},
\end{align}
where the prime means a derivative with respect to $z$, and 
$T^0_{\ 0}=T^z_{\ z}=-T^x_{\ x}=-T^y_{\ y}$ with
\be
T^0_{\ 0}=-\f{1}{2\mu_0}\left(E_0 e^{-b}\right)^2, \qquad W=-\f12 e^{a+2b}.
\ee

The field equations \eq{eint-eq1} implies $G^0_{\ 0}=G^z_{\ z}$, then $3b'^2+2b''=0$, whose solution is
\be \label{z2}
e^b=\lc 1+3 \lambda \lp z-z_*\rp \rc^{\f{2}{3}}\,,
\ee
where the constants $z_*$ and $\Ol$ have been written in the appropriated way to express invariance under translations and to get the limit $e^b\to 1$ when $E_0\to 0$ (here, $\Ol\to 0$).

We can now use \eq{z2} into \eq{gxx} to get $e^a$, but we prefer to use the constrain \eqref{gi00}, which leads to
\begin{align}
\Gamma^z_{00}&=-\f12 g^{zz}g_{00}'=\f12 e^{a+2b} \lp e^a\rp' \notag \\
&=\f{q}{m_0 c}e^{a/2}e^a \f{E_0}{c},
\end{align}
and by using \eq{z2}, we get
\be
e^a=\left(\f{q}{m_0c^2} \f{E_0}{\lambda}\right)^2 e^{-b}.
\ee

Considering that $e^a=e^b=1$ when $E_0\to 0$, then necessarily $\lambda=\pm q E_0/m_0 c^2$, 
and using $a+d=-2b$, the final result is
\be \label{sol-weyl}
e^a=e^d=e^{-b}=\lc 1+3 \lambda \lp z-z_*\rp \rc^{-\f{2}{3}}, \quad \lambda=\f{q E_0}{m_0 c^2}\,,
\ee
where we took the positive sign $\Ol$ as it can be always absorbed into $E_0$ by properly choosing the orientation of the $z$-axis as being either parallel or anti-parallel to the electric field.


\section{Analyzing the equations of motions}\label{ch:eqEx-mov} \label{ch:3-const}

In this section we examine the equations of motion in the \pst\ and their consequences on the dynamics of the particle. The treatment for the spherical case is very similar to the usual one of the  Schwarzschild metric, so we do not detail on this here. The case of \pst\ of a particle inside a constant electric field deserves more attention. From the previous section we have for $\vec{E}=E_0\hat z$ (we take without loosing generality $E_0>0$)
\begin{align} \label{metr}
ds^2&=\xi \lp -c^2dt^2+dz^2\rp +\xi^{-1}\left(dx^2+dy^2\right),\notag\\
&\xi=\lc 1+3\lambda \lp z-z_*\rp\rc^{-2/3}, \quad \Ol= \f{qE_0}{m_0 c^2}.
\end{align}

We see that the metric has a singularity at  $3\lambda (z-z_*)=-1$. This appears to be totally opposite to what we expect from translational invariance. However, we will now see that initial conditions do not allow the particle to reach such singular point.

As we will show later (see \eq{geod-z}), setting $v_z=0$ leads us to a quadratic equation in $\xi$ which gives two solutions, one with $\xi>0$ and other with $\xi<0$. 
The physical solution is that of positive $\xi$, then necessarily we have 
\be
\lambda \lp z_0-z_*\rp > -1\,,
\ee
where $z_0$ is the coordinate such that $v_z=0$.

\begin{figure}[!ht]
\begin{center}
         \includegraphics[width=.7\linewidth]{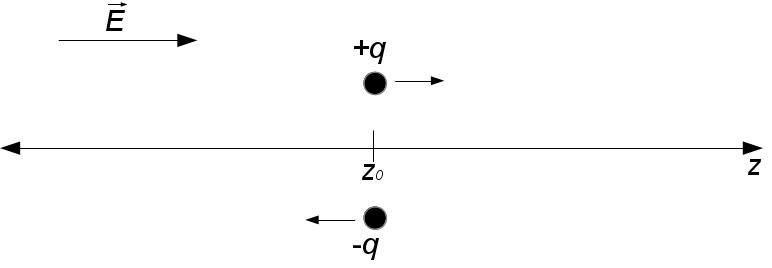}
\caption{Direction of motion from the return position $z_0$.}\label{fig:return}
\end{center}
\end{figure}

As we can see intuitively and is shown in Figure \ref{fig:return}, a positive charge ($\lambda>0$) will move to the right of $z_0$ due to the action of the electric field $E_0>0$, and a negative charge will move to the left. So, for $\Ol>0$ we have $z\ge z_0$, while for $\Ol<0$ we have $z\le z_0$ for all $z$ in the particle's trajectory, then we always have $\lambda \lp z-z_0\rp \ge 0$. As a consequence 
\be
\lambda \lp z-z_*\rp =\lambda \lp z-z_0\rp +\lambda \lp z_0-z_*\rp 
\ge \lambda \lp z_0-z_*\rp >-1\,,
\ee
and therefore the particle never reaches the singularity and in this way it respect the translational invariance. In fact, if we redefine $z \to \bar z=z-w$, then necessarily $z_0 \to \bar z_0=z_0-w$ and $z_* \to \bar z_*=z_*-w$; consequently  $\bar z-\bar z_*=z-z_*$.

If for example the particle starts at position $z_1$ and goes in direction of $z_0$, then when arriving at $z_0$ their velocity will change sign, forcing the particle to go back. We will then call $z_0$ the return position.

We know (see Carroll \cite{Carroll:2004st}) that if the metric does not depend on a 
given coordinate $x^\mu$, then the corresponding momentum
\be 
p_\mu=g_\Mn p^\nu=m g_\Mn \der{x^\nu}{\tau}\,,
\ee
is a constant of motion. So, the conserved quantities associated to the metric \eq{metr} are: $E=-p_0\ c/m$, $P=p_x/m$ and $p_y$. 

The normalization condition $g_\Mn (dx^\mu/d\tau) (dx^\nu/d\tau)=-1$ is (hereafter we set $p_y=0$)
%
\begin{align}
-1 &=\xi \lc -c^2 \lp \der{t}{\tau}\rp^2+\lp \der{z}{\tau}\rp^2\rc +\xi^{-1} \lp\der{x}{\tau}\rp^2\,,
\end{align}
and using the relation $E\f{d}{dt}=c^2 \xi\f{d}{d\tau}$, $v_z=dz/dt$, 
we get
\be \label{geod-z}
v_z^2=c^2-\f{P^2}{E^2}c^4\xi^2-\f{\xi}{E^2}c^6.
\ee

On the other hand $P=constant=P(z=z_*)$, then 
\be
P=\frac{1}{\xi} \der{x}{\tau}=\lp \f{E}{\xi^2 c^2} \f{dx}{dt}\rp \Big|_{z_*}=\f{E}{c^2} v_{x*}
\ee
where we define $v_{x*}$ and $v_{z*}$ as the $x$ and $z$ velocities at $z=z_*$. Now substituting into \eq{geod-z} yields
\be
v_{z*}^2=c^2-\lp v_{x*}\rp^2-\f{c^6}{E^2}, \quad \to \quad E=\f{c^2}{\sqrt{1-v_*^2/c^2}}\,,
\ee
with
\be
v_*^2=v_{z*}^2+v_{x*}^2.
\ee

We now concentrate on the case $P=0$, so $v_z=v$ and from \eq{geod-z} 
\be \label{eq-xi-v}
\xi=\lp 1-\f{v^2}{c^2}\rp \f{1}{1-v_*^2/c^2}, \quad \to \quad 1+3\lambda \lp z-z_*\rp=
\lp\f{1-\f{v^2}{c^2}}{1-v_*^2/c^2}\rp^{-3/2}\,,
\ee
from which the limit of small velocities  $v\ll c$ leads to
\be
\lambda \lp z-z_*\rp=\f{v^2}{2c^2}-\f{v^2_*}{2c^2}+{\mathcal O}(c^{-4}), \quad \to \quad 
\f12 m_0 \lp v^2-v^2_*\rp =qE_0 \lp z-z_*\rp+{\cal O}(c^{-2}),
\ee
which is just the energy conservation in the Newtonian limit ($\Delta E_{kinetic}=-\Delta E_{potential}$).

\subsection{Comparing with Special Relativity} \label{comparing}

In the SR, a charged particle in presence of an electric field $E_0$ satisfies the 
energy conservation (see \cite{GR5,GR6})
\begin{eqnarray}
qE_0 (z-z_*)=\f{m_0 c^2}{\sqrt{1-v^2/c^2}}{\Big|}^v_{v_*}\,,
\end{eqnarray}
which we write as
\be \label{prediction SR}
\lambda (z-z_*)=\f{1}{\sqrt{1-v^2/c^2}}-\f{1}{\sqrt{1-v^2_*/c^2}}\,.
\ee

To compare the two approaches, let's take, without loosing generality, $z_*=z_0=0$, 
so $v_*=0$ and then from equations (\ref{eq-xi-v}) and (\ref{prediction SR}) follow
\begin{align}
&\text{\pst:}\to \quad &\lp 1+3\lambda z\rp^{1/3}=\f{1}{\sqrt{1-v^2/c^2}} \label{pred-stp}\,, \\
&\text{SR:}\to \quad &\lp 1+\lambda z\rp=\f{1}{\sqrt{1-v^2/c^2}}\,. \label{pred-SR}
\end{align}

To have an idea of which order of magnitude in usual units we are talking about, consider the example given in Bouda and Belabbas \cite{Bouda:2010tb}. For a dust particle of mass $m\approx 10^{-14}kg$, saturation charge%
\footnote{Note that we are not in the quantum limit.} %
$q\approx 10^{-18}C$, and $c^2=9\times 10^{16}m^2/s^2$, we have
\be \label{campo-aprox}
\lambda \approx 10^{-21}E_0/V\,,
\ee
where $V$ means volts. An electric field as intense as $E_0\approx 10^8V/m$ is rare in practice (Bouda and Belabbas \cite{Bouda:2010tb}, see also \cite{Rafelski:2009fi}), for such an intense field we have $\lambda \approx 10^{-13}/m$ (here $m$ stands for meter). Then for such a particle moving in a region $|\Delta z|$ as big as a billion meter size 
the linear approximation works quite well and so \eq{pred-stp} coincides with \eq{pred-SR}.\\

\begin{figure}[!htb]
\includegraphics[width=.5\linewidth]{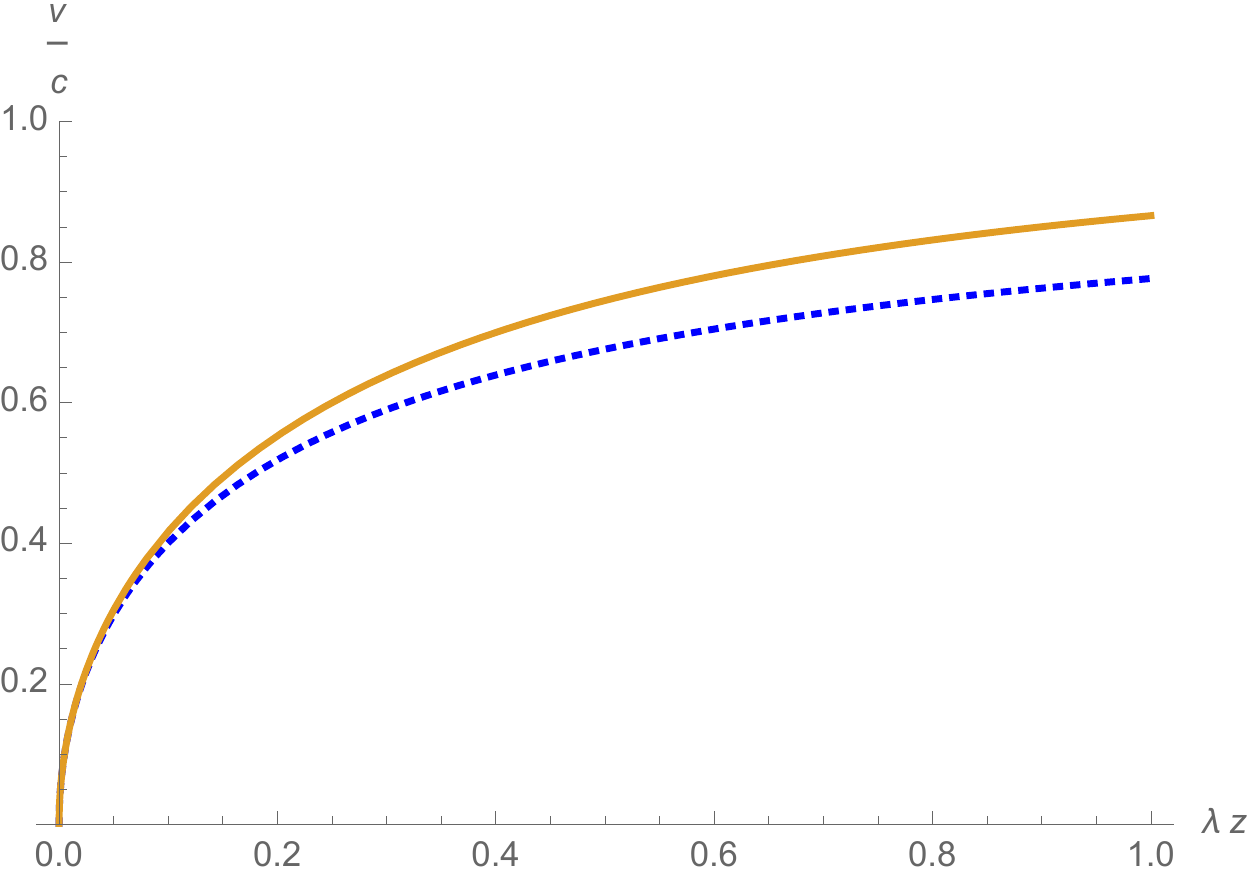}\quad
\includegraphics[width=.5\linewidth]{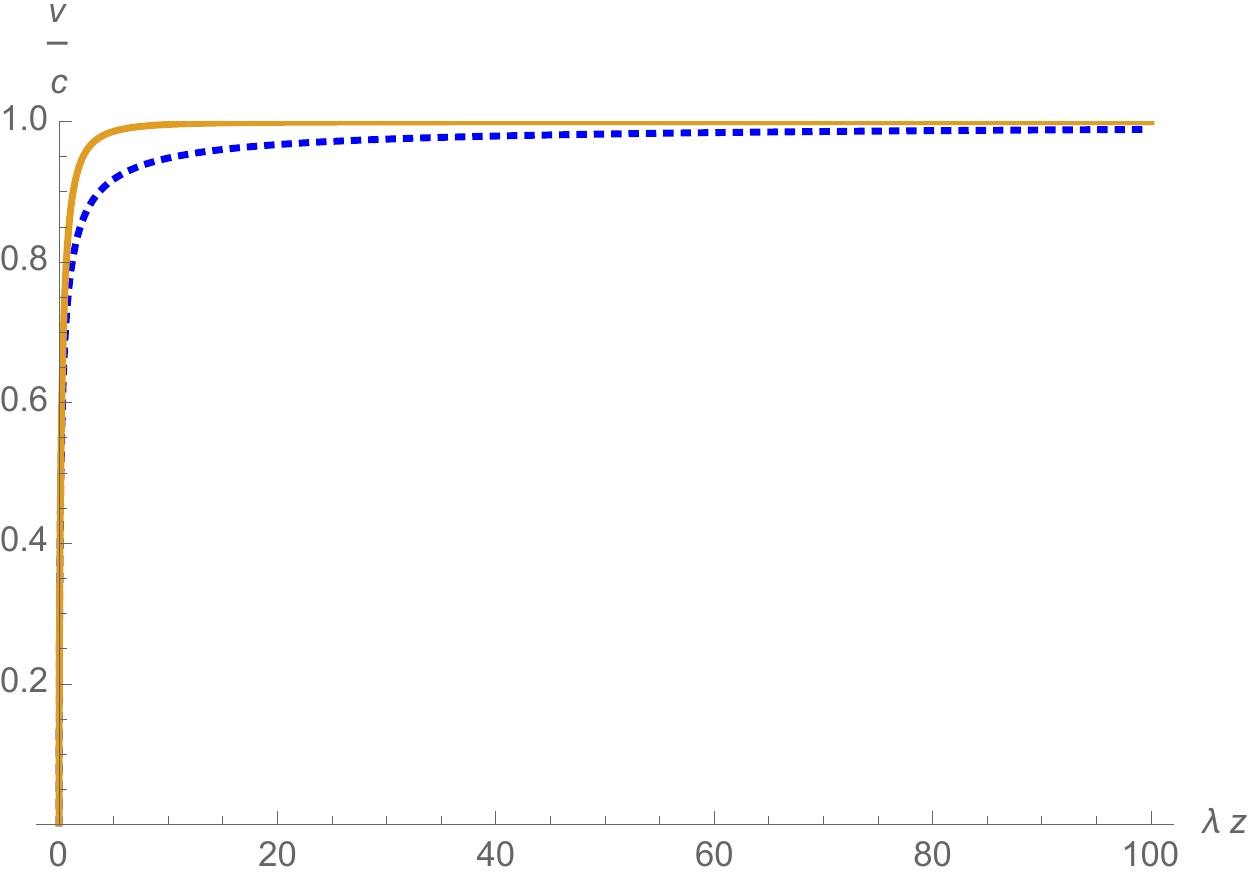}
\caption{Velocity vs position. Dotted lines are the prediction of \eq{pred-stp} 
while continuous lines are predicted by SR \eq{pred-SR}.}\label{fig:compare1}
\end{figure}

Figure \ref{fig:compare1} compares the two approaches, continuous lines are the SR solutions while dashed lines are the solutions for the \pst. For small values of $\lambda z$, say 
$\lambda z<0.1$ the two cases agree very well. We also see that in the \pst\ formalism, the particle approach the limit velocity $v=c$, slower than in the SR case. 

%

\subsection{Solution to the equation of motion}

In this section, in order to complete our analysis, we obtain the time dependence of the particle's path. For simplicity we restrict our calculations to the one dimensional case $P=0$, and again we take $z_*=z_0=0$ (or $v_*=0$).

In this case we have (\eq{pred-stp})
\begin{align} \label{eq-v-xi}
u^{-2/3}=1-v^2/c^2,\quad \to \quad cdt=\pm \f{dz}{\sqrt{1-u^{-2/3}}},
\end{align}
with $u=1+3\lambda z$. Then, in terms of $u$, \eq{eq-v-xi} yields
\be 
3\lambda cdt=\pm \f{u^{1/3} du}{u^{2/3}-1}, \quad \to \quad %
3\lambda c(t-t_0)=\pm \lp 2+u^{2/3}\rp \sqrt{u^{2/3}-1}\Bigg|^u_{u_0}
\ee
with $u_0$ the value of $u$ at $t=t_0$. Taking $t_0=0$, and using again the variable $z$ we get
\begin{align} \label{curv-geod}
(3\lambda)(ct)&=\pm \lp f(z)-f(z_0)\rp, \\
f(z)&=\lc 2+\lp 1+3\lambda z\rp^{2/3}\rc \sqrt{\lp 1+3\lambda z\rp^{2/3}-1}.
\end{align}
This equation gives the time as a function of the position. However we can isolate the variable $z$ as a function of time, and the reader can check that this leads to
\be
1+3\lambda z=\lc -1+\f{2^{2/3}}{2+s^2+s\sqrt{4+s^2}}
+\f{2+s^2+s\sqrt{4+s^2}}{2^{2/3}}\rc^{3/2}\,,
\ee
with
\begin{align}
s=\pm (3\lambda)(ct)+f(z_0).\label{s}
\end{align}

\begin{figure}[!ht]
\centering
\includegraphics[width=.7\linewidth]{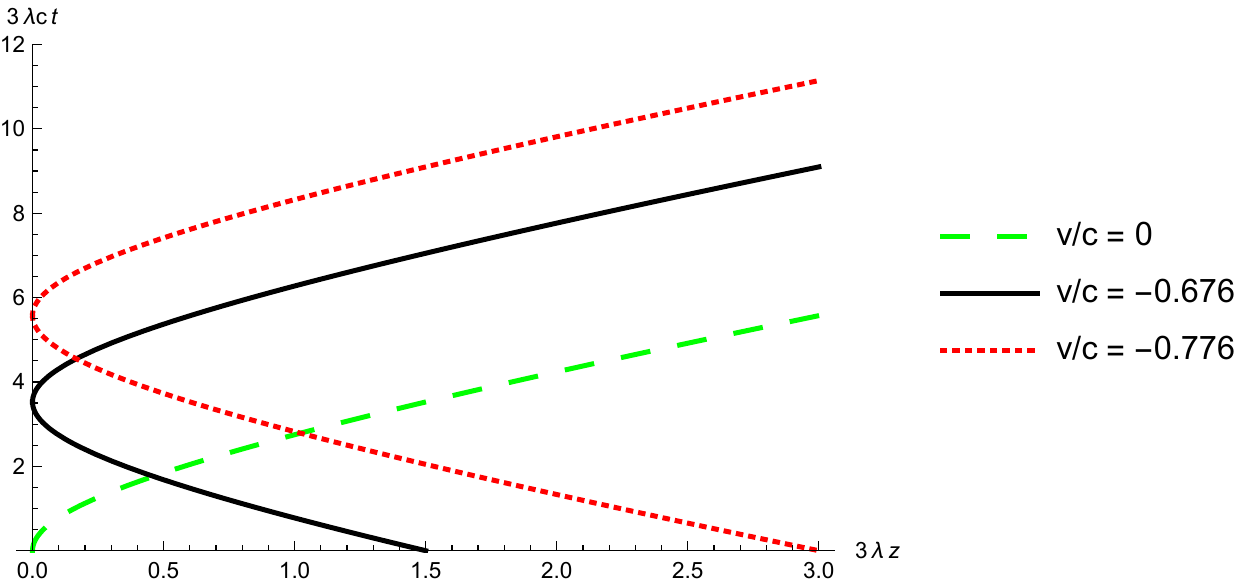}
\caption{Geodesic curves for the one-dimensional case. Initial conditions shown on the right.}\label{fig:geodesics}
\end{figure}

Figure \ref{fig:geodesics} shows a set of geodesics for different initial conditions: $3\lambda z=0,\ 1.5,\ 3$. 
The cases $3\lambda z=1.5$ and $3\lambda z=3$, start with negative velocity and approach to the return point (in this case, the origin), then go back to the $z>0$ region. The other case is for initial conditions $3\lambda z=0$ and $v=0$, so they are always moving away from the origin.

The initial velocities for each case, can be obtained from \eq{pred-stp}
\be
\f{v}{c}=\pm \sqrt{1-\lp 1+3\lambda z\rp^{-2/3}},
\ee
which yields
\begin{align}
3\lambda z=1.5 \quad &\to \quad 	\f{v}{c} 	\approx 0.676, \\
3\lambda z=3 \quad &\to \quad	\f{v}{c} 	\approx 0.776\,.
\end{align}

In this plot we explicitly see that due to the initial conditions and the return position, the particle 
never approach the region $\lambda z<0$ (here $z_*=0)$, this makes the \pst\ described by \eq{metr} free of 
singularities. We see also that basically each curve is equal to the other up to a displacement, this agrees with translational invariance as we would expect.


\section*{Summary and Conclusions}

In this work we have introduced the concept of \pst\ as a geometric tool to describe the interaction between a punctual particle and an electromagnetic external field. In this formalism we have considered the particle path as being a geodesic in a curved space-time, and so, the electromagnetic force could be understood in a purely geometric way. 

With those considerations we have seen in section \ref{lorentz} that the geodesic equation in the \pst\ leads to the Lorentz force in the limit of weak field and small velocities. This allowed us to relate the metric in the \pst\ with the EM potential. In section \ref{ch:campoFmn} we have proposed general statements about the characteristics a the \pst\ beyond the weak field approximation which allowed us to make some applications in section \ref{sec:appl}. 
In particular, we have obtained the field equations that define this geometry which are similar to the Einstein's equations, where in general, the energy-momentum tensor have information of both, the particle and the external field. We have seen however, that those field equations violate gauge invariance, as they are only valid in a specific gauge. This is a question that must be studied more carefully and will be left for future works. We have also confirmed that the present treatment is compatible with SR in the limit of small velocities and weak fields (see section \ref{comparing}), that means that the general behavior of the particle has been obtained. An interesting result is that it is possible to find connexions to the usual formulation, and the conditions in order to observe these connections are determined. Thinking in these terms, the proposed theory may be seen as a generalization of the usual theory.

\paragraph{Acknowledgments} We thank Brazilian research agency CNPq for financial support.


\end{document}